\documentclass[aps,prc,twocolumn,floatfix,showpacs,a4paper,
nofootinbib,amsmath,amssymb]{revtex4-1}
\usepackage[colorlinks=true,linktocpage=true,linkcolor=blue,citecolor=blue,allcolors=blue]{hyperref}
\usepackage{graphicx}
\usepackage{dcolumn}
\usepackage{bm}
\usepackage{color}
\usepackage{amssymb}
\usepackage{amsthm}

\newcommand{\be}{\begin{equation}}
\newcommand{\ee}{\end{equation}}
\newcommand{\ba}{\begin{eqnarray}}
\newcommand{\ea}{\end{eqnarray}}
\newcommand{\bd}{\begin{displaymath}}
\newcommand{\ed}{\end{displaymath}}
\newcommand{\bea}{\begin{eqnarray}}
\newcommand{\eea}{\end{eqnarray}}

\newcommand{\hsp}{\hspace*{0.5pt}}
\newcommand{\hsl}{\hspace*{1pt}}

\begin{document}

\title{Crater Formation and Deuterium Production in
\\Laser Irradiation of Polymers with Implanted Nano-antennas}

\author{
L\'aszl\'o P. Csernai$^{1,2,3}$,
Igor N. Mishustin$^{3}$,
Leonid M. Satarov$^3$,
Horst St\"ocker$^{3,7,8}$,
Larissa Bravina$^4$,
M\'aria Csete$^{5,6}$,
Judit K\'am\'an$^{1,5}$,
Archana Kumari$^{1,5}$,
Anton Motornenko$^3$,
Istv\'an Papp$^{1,5}$,
P\'eter R\'acz$^{1,5}$,
Daniel D. Strottman$^9$,\\
Andr\'as Szenes$^{5,6}$,
\'Agnes Szokol$^{1,5}$,
D\'avid Vass$^{5,6}$,
Mikl\'os Veres$^{1,5}$,
Tam\'as S. Bir\'o$^{1,5}$,
Norbert Kro\'o$^{1,5,10}$\\
(NAPLIFE Collaboration)\\
$  $}

\address{
$^1$Wigner Research Centre for Physics, Budapest, Hungary\\
$^2$Deptartment of Physics and Technology, University of Bergen, Norway\\
$^3$Frankfurt Institute for Advanced Studies, Frankfurt am Main, Germany\\
$^4$Department of Physics, University of Oslo, Norway\\
$^5$National Research, Development and Innovation Office of Hungary,\\
$^6$Deptartment of Optics and Quantum Electronics, Univ. of Szeged, Hungary\\
$^7$Institute f\"ur Theoretische Physik, Goethe Universit\"at, Frankfurt am Main, Germany\\
$^8$GSI Helmholtzzentrum f\"ur Schwerionenforschung GmbH, Darmstadt, Germany\\
$^{9}$Los Alamos National Laboratory, Los Alamos, 87545 NM, USA \\
$^{10}$Hungarian Academy of Sciences, 1051 Budapest, Hungary
}

\begin{abstract}
Recent validation experiments on laser irradiation of polymer foils with
and without implanted golden nano-particles are discussed.
First we analyze characteristics of craters, formed in the target after
its interaction with laser beam.
Preliminary experimental results show significant production
of deuterons when both the energy of laser pulse and concentration
of nano-particles are high enough. We consider the deuteron production
via the nuclear transmutation reactions $p+C\hsp\rightarrow\hsp d+X$
where protons are accelerated by Coulomb field, generated in the target
plasma. We argue that maximal proton energy can be above threshold values
for these reactions and the deuteron yield
may noticeably increase due to presence of nano-particles.
\end{abstract}

\maketitle{}
.

\section{Introduction}

Relativistic heavy-ion collisions have shown that
the hadronization (or burning)  {of created matter} is a very fast, nearly
simultaneous process in proper-time. This is in contrast to the
classical Rayleigh-Hugoniot-Taub detonation theory.
The possibility of relativistic detonation on a
time-like hyper-surface was  {theoretically discovered} in 1987
\cite{Cs1987},
and  {was} later applied to  {simulate} pellet fusion in
 {Ref.}~\cite{CS2015}.
 {It was found that it is not possible to obtain} a better result than
 {was reached in NIF experiments}, where Rayleigh-Taylor instabilities
prevented the full burning of the target fuel.

To remedy this problem, we have suggested~
\cite{CKP2017p}
to implant golden nano-particles (nano-antennas) into the target.
In this way one can achieve "time-like"
ignition, which occurs nearly simultaneously in the whole target volume.
Below we describe the validation experiments within the
NAnoPlasmonic Laser Induced Fusion Energy (NAPLIFE) project
\cite{CsEA2020}.  {They are}
planned in two steps: (i) verification of the
amplified absorption of the laser energy  {in}
the target  {with nano-particles}, and
(ii)  testing a simultaneous detonation in the whole
target volume, induced by two short ($\tau\lesssim 1~\textrm{fs}$)
laser pulses
radiating opposite sides of a flat target\!
\footnote{The two-sided irradiation was
applied earlier in~Refs.~\cite{Bonasera2019,ZhangJ-DCI-2020}.}.

For the validation tests we have chosen
the polymer UDMA  (see below), which is hard enough to make thin (3\,$\mu$m) layers
of it and construct an altogether 21\,$\mu$m multi-layer target.
 {By adjusting the nano-particle~(NP) density profile one can achieve}
\cite{Csete2021}
simultaneous  {light absorption} in the whole volume.

Recently we started
validation experiments where a~thicker ($\sim 160\,\mu$m)
target was irradiated from one side leading to the formation
of a crater \cite{Racz-22,Kroo-22}.
\bigskip
\bigskip

\section{Crater formation}
\subsection*{a)~Composition of the target}

The Urethane Di-MethAcrylate (UDMA) polymer contains 470 nucleons
and 254 electrons in 71 atoms. Its molecular formula is
\mbox{C$_{23}$H$_{38}$N$_2$O$_8$}.
 {The calculated mass of this molecule equals
$470.2628~\textrm{amu}$~\cite{udma}.}
The reference atom, $^{12\hsp}$C has in vacuum the binding energy of
7.68 MeV/nucleon. Compared to this value the average
binding energy per nucleon in the UDMA is  {equal}
$0.2628/470~\textrm{amu}\simeq 0.5207~\textrm{MeV}$.

After the irradiation  {by} an energetic laser pulse the
UDMA molecule may fragment and break up to pieces.
 {In~this process} numerous chemical configurations, e.g.,
2\hsp (NO$_2$),\hsl 2\hsp (CO$_2$),\hsl CH$_4$,\hsl C$_{16}$H$_{34}$,\hsl 4\hsp C\hsp,
or even  {single atoms may appear in the final state. At~large enough
laser} energies, recombination of
nucleons becomes also possible,  {and new atomic species can be
formed. For example,} a~neutron from $^{13\hsp}$C may be transferred to
a~Hydrogen,  {producing} a~Deuterium.  {This is an endothermic
reaction, which requires} more energy than  {needed for
the molecular} fragmentation.

As far as we know, the Deuterium formation in laser irradiated targets
was observed in these validation experiments for the first time.
It  is interesting that its amount strongly  increases with raising
density of implanted golden nano-antennas
\cite{Kroo-22}.

In our validation experiments, we have irradiated three types of
thick targets: pure UDMA without nano-antennas (Au0), as well as
UDMA with  {golden} nano-rod antennas of  {the} size 25x85\,nm and
volume
\mbox{$V_{Au}\simeq 41724\,{\rm nm}^3$}.
{We have used targets with two different}
mass density ratios  {of NPs}:
$ m_{Au}/m_{U}=0.126\%$ (Au1) and $0.182\%$~(Au2).
It was observed that laser shots brake out craters in the UDMA target.
Parameters of these crater were analyzed by microscope.
It was found that they depend significantly on
the concentration of NPs.

\subsection*{b)~Craters with nano-plasmonics} 

 Geometrical parameters of craters
were measured  in
Refs.~\cite{Szokol-NFP22,VeresPP,Kaman2021}.
The laser beam with a Gaussian intensity profile
was focused to a spot with diameter of 24.5$ \pm 3.2 \mu$m up to 85\% norm
(FWHM $ \simeq 14.4\,\mu$m, the shape was not circularly
symmetric)~\cite{Szokol1}.

\begin{figure}[h]  
\begin{center}
\includegraphics[width=0.99\columnwidth]{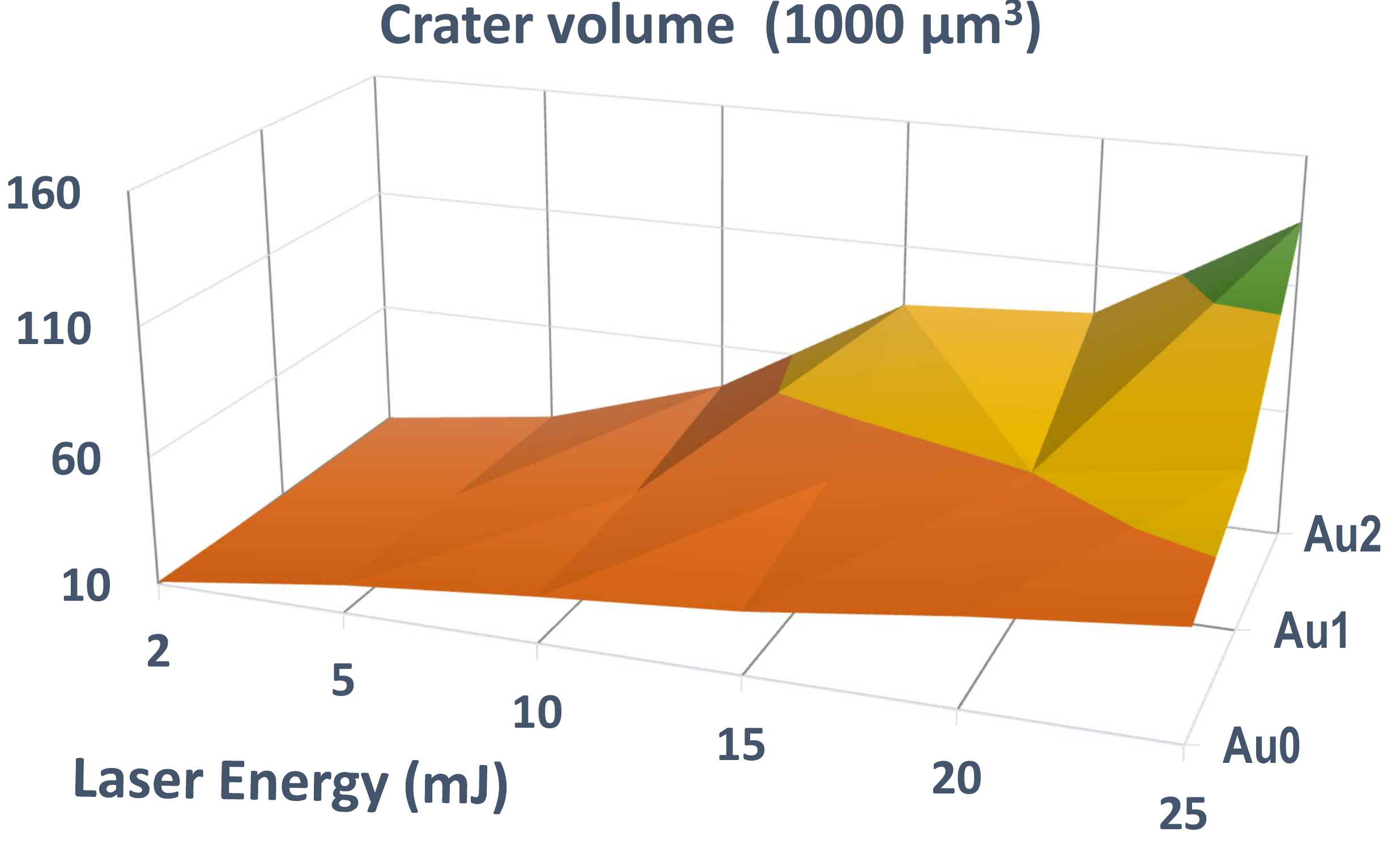}
\end{center}
\vskip -5 mm
\caption{
(color online)
The crater volume $V_{cr}$ in the thick UDMA target
(in units of 1000 $\mu\hsp\textrm{m}^3$)
as the function of laser pulse energy $E_L$. Several types
of targets are  considered: Au1 and Au2 with implanted nano-rod antennas,
and Au0 without implantation.
The mass concentrations of implanted particles in UDMA
are 0.126\%
and 0.182\%
for targets Au1 and Au2, respectively.
For each target type five pulse energies are considered.
At the highest pulse energy the crater volume increases rapidly
with increasing implantation density.
For more details, see Refs.~\cite{Szokol-NFP22,Kroo-22}.
}
\label{F-0}
\end{figure}

\begin{figure}[h]  
\begin{center}
\includegraphics[width=0.99\columnwidth]{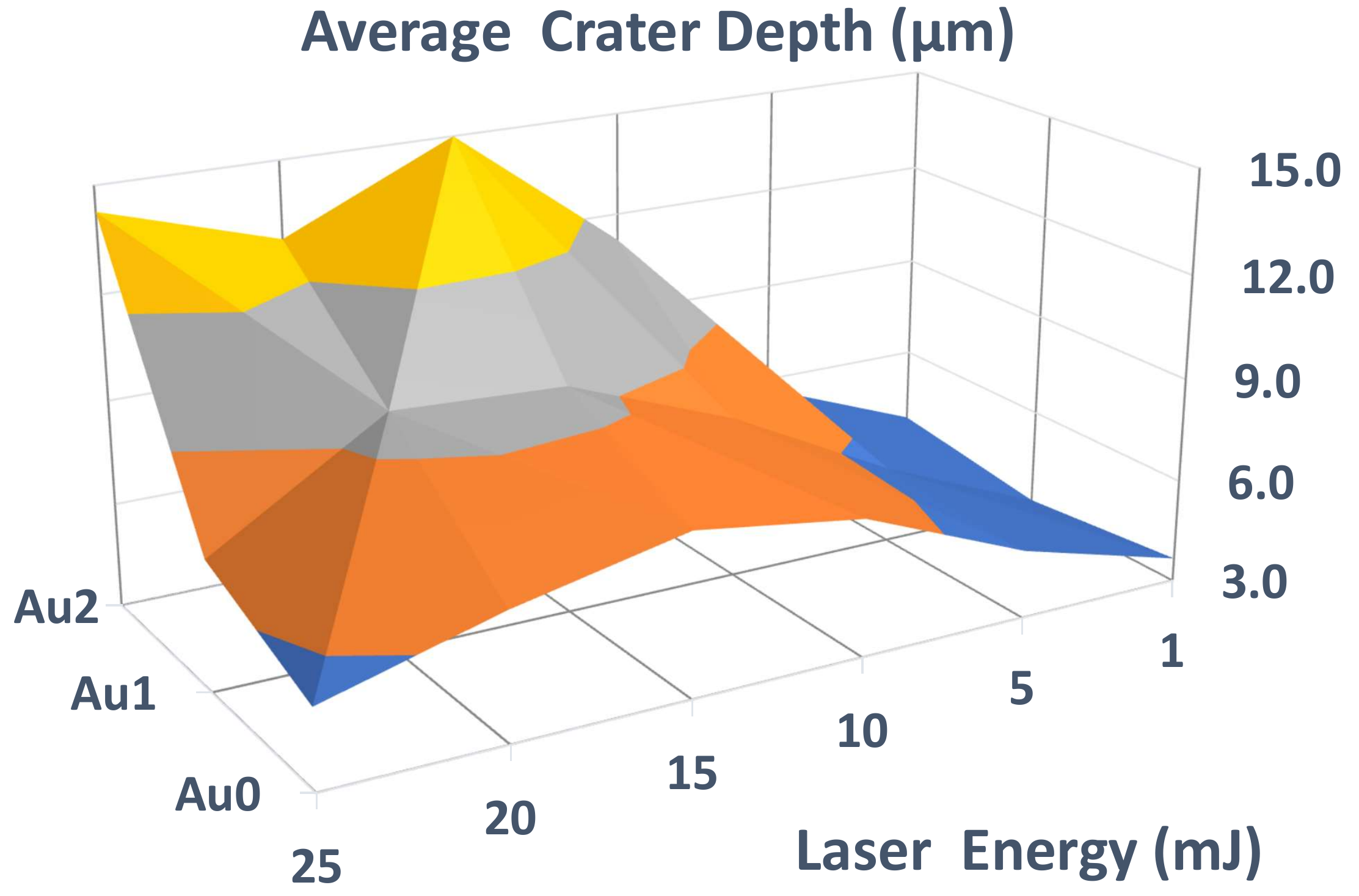}
\end{center}
\vskip -4 mm
\caption{
(color online)
Average crater depth assuming a circular cone shape as
a~function of the energy of the laser irradiation~$E_L$ and the
type of target material. The depth is minimal (3.7--7.7\,$\mu$m)
without nano-rod antennas~(Au0), while for Au2 target
it increases from 5.7 to 15.6\,$\mu$m.
}
\label{F-3}
\end{figure}

The following values of crater {diameters} have been obtained:
110--130 $\mu$m at the laser pulse energy $E_L=5~\textrm{mJ}$,
120--130 $\mu$m at $E_L=10~\textrm{mJ}$,
130--150 $\mu$m at $E_L=15~\textrm{mJ}$,
155--165 $\mu$m at $E_L=20~\textrm{mJ}$, and
190--200 $\mu$m at $E_L=25~\textrm{mJ}$.
The average crater depths, calculated from observed crater volumes and diameters
(assuming its cone shape), are as follows:
  5.0, 5.2,  5.7 $\mu$m at $E_L=5~\textrm{mJ}$,
  7.7, 9.6, 15.6 $\mu$m at $E_L=15~\textrm{mJ}$, and
  5.6, 6.7, 14.3 $\mu$m at $E_L=25~\textrm{mJ}$,
for the targets Au0, Au1, Au2, respectively.
As the effective focus of the laser beam was fixed to
about {12.6} $\mu$m\,,
the observed increase of the crater diameter could be attributed to
the transverse expansion of hot plasma created by the laser beam.
One can see that the crater diameters do not depend much on the
concentration of nano-antennas. On the other hand, the
crater volumes and depths clearly depend on it, see
Figs.~\ref{F-0},~\ref{F-3}.

One can see  in Fig.~1 that the crater volume $V_{cr}$
increases with the  laser pulse energy and with the density of
nano-antennas.  As compared to the pure UDMA target (Au0),
the crater volume increases  with $E_L$ more rapidly
for targets with nano-antennas (Au1 and Au2).
Changing the nano-rod antenna
density from zero (Au0) to~0.182~\% (Au2) increases the crater volume
from 50000 to 135000 $\mu$m$^3$. Thus, the nano-antennas increase
the laser light absorption.
The largest crater volume
\mbox{$V_{cr}=135000~\mu\textrm{m}^3$}
was observed in the experiment Au2 at
\mbox{$E_L=25~\textrm{mJ}$}.

It is interesting to note that the crater volume increases faster than
the deposited laser energy. This indicates that the
energy of excavating the crater is
appreciably larger  than the energy directly coming from the
laser. An additional energy may come from
exothermic chemical reactions during the
fragmentation of UDMA, which are neglected in the present
analysis.

The average distance between nano-antennas in the unperturbed
target can be estimated as
\be
\langle L_{AuX} \rangle {=}
\left[V_{Au} (\rho_{Au}/\rho_{U}) (m_{U}/m_{AuX}) \right]^{1/3},
\label{LA}
\ee
where $V_{Au}$ and \mbox{$\rho_{Au}\simeq 19.32~\textrm{g/cm}^3$}
are, respectively,  the volume and mass density of a~single
NP, and \mbox{$\rho_U\simeq 1.14~\textrm{g/cm}^3$}
is the mass density of a pure UDMA material.
Using Eq.~(\ref{LA}), one gets the values
\mbox{$\langle L_{Au1} \rangle = 0.82~\mu{\rm m} $} and
\mbox{$\langle L_{Au2} \rangle = 0.73~\mu{\rm m} $}.
This relatively sparse implantation is well visible on the
image of the irradiated Au1 and Au2 targets
\cite{Kaman2021}.
In the crater image one can clearly see irregular "cracks"
associated with nano-rod positions~\cite{VeresPP}.
We think that these cracks are formed after 'explosion' of
nano-rods located near the bottom of the crater.

\subsection*{c)~Energy densities in target}  

\begin{figure}[h]  
\begin{center}
\includegraphics[width=0.99\columnwidth]{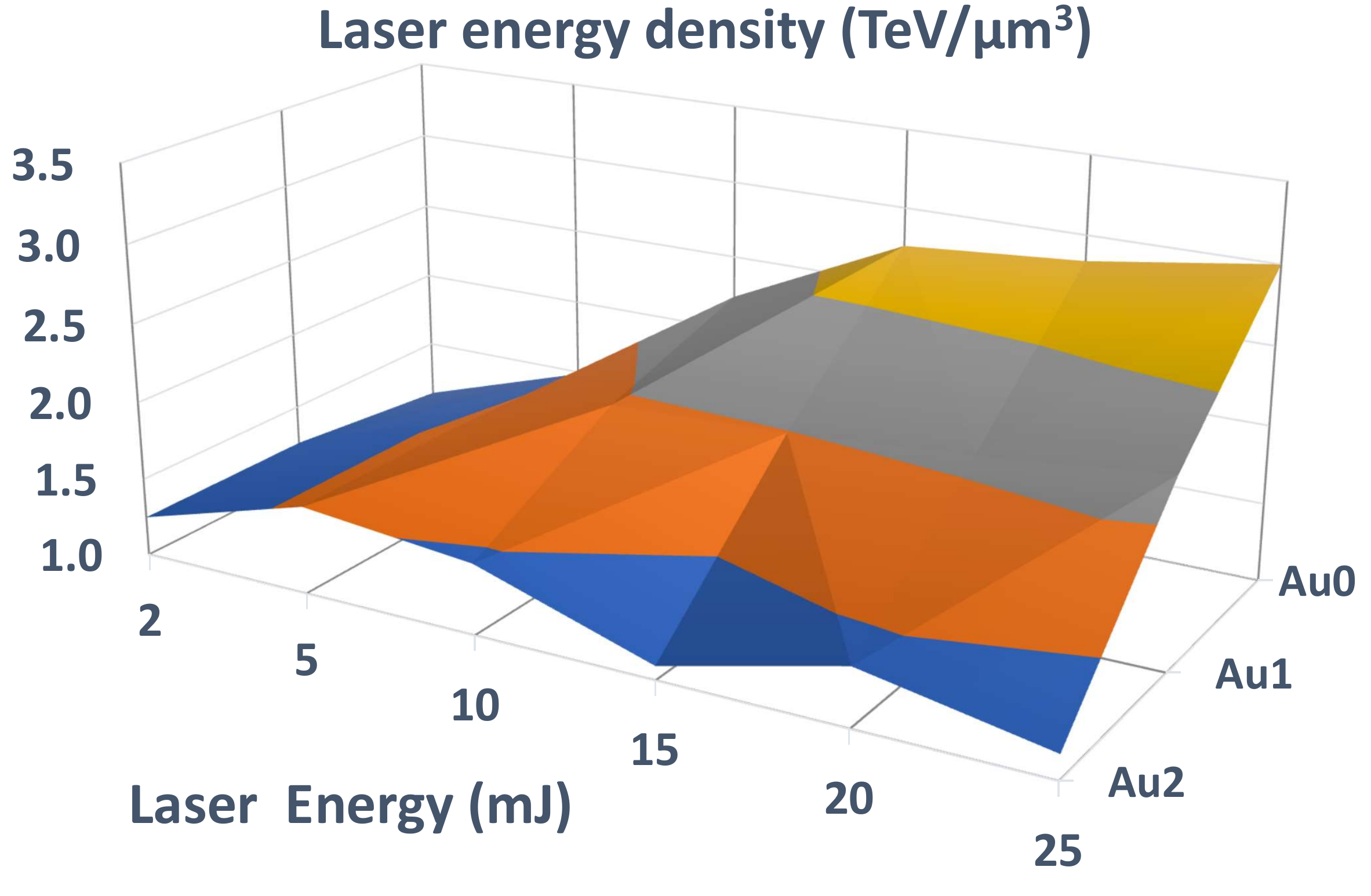}
\end{center}
\vskip -9mm
\caption{
(color online)
The laser pulse energy divided by the volume of the crater
$E_L/V_{cr}$.
\mbox{Average} energy density in the crater of the target is not
directly measurable. In the Au2 target
the laser energy is mostly absorbed
in the crater, while in two other cases a~larger part of the
laser energy is passed through or reflected out of it. So more
laser irradiation energy is needed to produce the same crater size.
}
\label{F-2}
\end{figure}

One can roughly estimate the average energy density deposited in the target
as \mbox{$\varepsilon_{cr}=E_L/V_{cr}$}.
This gives the results shown in Fig.~\ref{F-2}.
One can see that
$\varepsilon_{cr}\sim 1~\textrm{TeV}/\mu\hsp\textrm{m}^3$
for the top laser energy and NP density.
The energy density from the laser beam at higher {$E_L$} and
implantation density decreases,
because the crater volume increases more rapidly\hsp\footnote
{
It is clear
that the energy of the laser beam should not be
equal to the energy absorbed by the target matter, since a part
of the beam is reflected.
}.
This effect indicates that the nano-rods cause an~additional
absorption of laser energy, this is why we call them antennas.


In case of $E_L=25~\textrm{mJ}$ and the Au2 target
one can estimate the energy  absorbed by a single NP as
\mbox{$\epsilon_{cr}\hsp V_{Au}\sim 50~\textrm{MeV}$}, where $V_{Au}$
is the volume of a single nano-rod.
In this case the initial number of nano-antennas in the  crater
volume is about $V_{cr}/\langle L_{Au2} \rangle^3\simeq 347000$.

Note, however, that these energy estimates are very crude.
In fact, interaction of a~short laser pulse with a~dense target
is a~very complicated, space-time dependent process.
Presumably, at an early stage the laser beam excites
a target in a focal cylindrical channel, which is much more narrow
compared to a crater region formed at late times. A~part of
laser energy goes to ionization of target's atoms, leading
to formation of electron-ion plasma. Under influence of
electric fields
and microscopic collisions the target particles accelerate
and some of them may escape from the crater region.
The plasmon excitations in nano-antennas may create
'hot spots' in surrounding matter
with enhanced energy density~\cite{Papp2022}.

\section{Laser-induced deuterium production}
\subsection*{a) Experimental results}

As a result of laser  irradiation, many electrons
as well as molecular and atomic ions are formed in
the target.  To our surprise, in the experiments with implanted
NPs we have observed~\cite{Kroo-22,Kumari-22}
a significant fraction of Deuterium atoms, which are
emitted on the level of up to $10\%$ of the observed Hydrogen numbers,
D/(2\hsp D+H). Here D and H are the {\it observed}
yields of Deuterium and Hydrogen \textit{atoms}, both
determined from the LIBS spectra~\cite{Racz-22},
measured at 0.5 microsecond delay time after the laser pulse~
\cite{Kumari-22}.
These atoms were detected by presence of
\mbox{Balmer-$\alpha$} lines
in the radiation of the H and D, at
$\lambda =$ 656.27~nm  and 656.11~nm,
($E_{B\alpha} =$ 0.944~623 eV  and 0.944~843 eV), respectively.
The absolute amounts of {\it observed} H and D were not measured.
The relative amount of Deuterium atoms increases linearly with the
laser beam energy, reaching 7\% of Hydrogen at $E_L= 25$\,mJ with a
target including nano-rod antennas with mass density of 0.126\% (Au1).

We have checked that originally the polymer foils used in our
experiments did not contain Deuterium. Below we analyze possible
mechanisms of D production.

\subsection*{b)~Nuclear transmutation reactions} 

Let us first assume that D atoms appear due to
{electron capture}
(inverse $\beta$~decay)
\be
p+e  \rightarrow n + \nu\,,
\label{weak}
\ee
followed by the $n+p\to d$ fusion
~\cite{elcon}.
The reaction~(\ref{weak}) is endothermic, and may occur
if the kinetic energy of the electron
exceeds 783~keV in the lab frame.
The fusion of $n$ and $p$ is exothermic process,
\be
n+p \rightarrow d + 2.225\,\textrm{MeV}\,.
\label{fu}
\ee
Note, that this reaction requires
an additional (spectator) particle.
However, the reaction (\ref{weak}) is a weak process
with a very small cross section. Therefore, this mechanism cannot explain
the observed yield of Deuterium~\cite{pptod}.

Below we focus on one important aspect of the UDMA fragmentation
process, which may lead to nuclear transmutation reactions.
Especially interesting
are the reactions associated with carbon atoms in the UDMA molecule.
Note that, a natural carbon is the mixture of
two stable isotopes: 98.4\%
of $^{12\hsp}$C and 1.06\% of~$^{13}$C.
Fast protons accelerated by laser irradiation may
lead to production of deuterons via the stripping reactions
\be
p + \mbox{$^{13\hsp}\textrm{C}$}\hsp\rightarrow\hsp d
+ \mbox{$^{12\hsp}\textrm{C}$}\,,
\label{pd}
\ee
and
\be
p + \mbox{$^{12\hsp}\textrm{C}$}\hsp\rightarrow\hsp d
+ \mbox{$^{11\hsp}\textrm{C}$}\,.
\label{pd1}
\ee
These reactions are endothermic
and may occur if the proton energy exceeds some minimal (threshold)
value.
The last neutron in the ground state of the isotope $^{13}$C
is relatively weakly bound,  with the
binding energy of about 4.95~MeV. On the other hand,
the neutron separation energy of $^{12}$C
equals approximately 15.65~MeV. Due to this reason, the threshold
proton energy required for the reaction (\ref{pd1}) is larger
than for the reaction (\ref{pd}) (see below).

Let us first consider reaction~(\ref{pd}). It has
a negative $Q$ value\hsp\footnote
{
Here we use numerical values of atomic masses from Ref.~\cite{AME2020}.
}
\begin{eqnarray}
Q\,&&= m\hsp ({^{13\hsp}\textrm{C}})+m_p-m\hsp
({^{12\hsp}\textrm{C}})-m_d\nonumber\\
&&\simeq -2.7216\,{\rm MeV}\, .
\label{QC}
\end{eqnarray}
Therefore, it can only take place  if protons
created during the laser-target interaction
have sufficiently high kinetic energies (see below).

The inverse 'pick-up' reaction
\be
d+\mbox{$^{12\hsp}\textrm{C}$}\hsp\rightarrow\hsp
p+\mbox{$^{13\hsp}\textrm{C}$}\,.
\label{dp}
\ee
is exothermic. It was extensively studied experimentally
in the past
(see, e.g., Refs.~\cite{Phi1950,Kas1960,Hod1967}).

The ratio of the cross sections for the pick-up and stripping reactions
can be calculated by using the detailed balance relation~\cite{Lan77}:
\be
\frac{\sigma_s}{\sigma_p}=\frac{3}{4}\left(\frac{p_d^{\,\rm cm} }
{p_p^{\,\rm cm}}\right)^2,
\label{dbr}
\ee
where $\sigma_s$ and~$\sigma_p$ are, respectively,
the cross sections of the reactions
(\ref{pd}) and~(\ref{dp}), $p_p^{\,\rm cm} (p_d^{\,\rm cm})$
is the c.m.~momentum of the proton (deuteron). All these
quantities should be taken at the same c.m. total energy $\sqrt{s}$.
The numerical coefficient
in the right hand side of Eq.~(\ref{dbr}) is the ratio of
spin degeneracy weights of particles in the final and initial
states of the reaction (\ref{pd}).

 Let us further apply the relations of relativistic kinematics
\vspace*{-2mm}
\begin{eqnarray}
s&&=2\hsp m\,(^{13\hsp}\textrm{C})\hsp E_p^{\,\rm lab}+\left[m_p+m\,
(^{13\hsp}\textrm{C})\right]^2\nonumber\\
&&=2\hsp m\,(^{12\hsp}\textrm{C})\hsp E_d^{\,\rm lab}+\left[m_d+m\,
(^{12\hsp}\textrm{C})\right]^2,
\label{rkin}
\end{eqnarray}
where $E_i^{\,\rm lab}$ is the kinetic energy of
particles $i=p,d$ in the lab frame.
The threshold proton energy for the reaction (\ref{pd}) is obtained
from Eq.~(\ref{rkin}) by substituting \mbox{$E_d^{\,\rm lab}=0$}.
Then we obtain that $\sigma_s\ne 0$ at $E_p^{\,\rm lab}$
above $2.934~\textrm{MeV}$, which is sightly larger
than \mbox{$|Q|$-value} determined in Eq.~(\ref{QC}).
Analogous calculation for the
reaction (\ref{pd1}) shows that the latter has the threshold
proton energy 14.568~MeV.

To calculate the ratio $\sigma_s/\sigma_p$
one can use the kinematic relation
\be
\frac{p_d^{\,\rm cm} }{p_p^{\,\rm cm}}=\frac{p_d^{\,\rm lab}}{p_p^{\,\rm lab}}
\cdot \frac{m\,(^{12\hsp}\textrm{C})}{m\,(^{13\hsp}\textrm{C})}\simeq
\sqrt{\frac{m_d\hsp E_d^{\,\rm lab}}{m_p\hsp E_p^{\,\rm lab}}}\,
\frac{m\,(^{12\hsp}\textrm{C})}{m\,(^{13\hsp}\textrm{C})}\,.
\label{pra}
\ee
Here the last equality is written in the non-relativistic approximation.
As an example, let us consider the deuteron production with energy
\mbox{$E_d^{\rm lab}=1~\textrm{MeV}$}
in the stripping reaction (\ref{pd}).
According to
Ref.~\cite{Phi1950}
the inverse reaction (\ref{dp}) has the cross section
\mbox{$\sigma_p\simeq 6~\textrm{mb}$}
at this bombarding energy. From Eq.~(\ref{rkin})
one can calculate the corresponding energy of
protons in the reaction~(\ref{pd}),
which equals $E_p^{\rm lab}=3.857~\textrm{MeV}$. \mbox{Using}
further Eq.~(\ref{pra}), one obtains the value
\mbox{$\sigma_s\simeq 0.330\,\sigma_p\simeq 2~\textrm{mb}$.}

Note, that many C-containing groups are present in the UDMA
molecules. E.g. the Methyl group $CH_3$ is frequent
at the sides and ends of the UDMA molecule.
The Methyl group is bounded to the rest of the molecule
by a single covalent bond ($- CH_3$).
In these groups the transition of a neutron
from $C$ to the $H$ is especially easy,
occurring via the reaction
\ba
\mbox{$^{13\hsp}\textrm{C}$}\textrm{H}_3 &\rightarrow&
(\mbox{$^{12\hsp}\textrm{C}$}{+}n)(\textrm{H}_2{+}p) \ \rightarrow\
\nonumber\\
\mbox{$^{12\hsp}\textrm{C}$}\textrm{H}_2 {+} (p{+}n) &\rightarrow&
\mbox{$^{12\hsp}\textrm{C}$}\textrm{H}_2 {+} D .
\label{strip}
\ea
In vacuum this reaction is also endothermic requiring
2.85 MeV additional energy.
(If doubled it is
$2\cdot(CH_3 + H)\ \rightarrow\  C_2H_4 +2D$,
where $C_2H_4$ is Ethylene).
Similar transmutation is possible with the
Methylene group $CH_2$. This group
is connected to the remainder
of the molecule by two single bonds
$-  CH_2 -$ (Methylene Bridge)  or $ CH_2\! =$.
The same reaction with the Methylene group is
producing $D$ and $C_2H_2$ (Acetylene).
Observing Ethylene or Acetylene as byproducts after the
Laser pulse irradiation, would confirm this mechanism
of the intra-molecular nucleon exchange.

\subsection*{c)~Laser-induced proton acceleration} 

It is well known that energetic
laser beams may strip electrons,
which then pull protons from the target
and accelerate them to multi-MeV
energy~\cite{PHELIX100,Rot2017,LWFC2021}.
In this section we argue that parameters of the NAPLIFE laser pulse
may be sufficient to drive the
$\mbox{$^{13\hsp}\textrm{C}$}\hsp (p,d)
\mbox{$^{12\hsp}\textrm{C}$}$
reaction in the UDMA target and, therefore,
to contribute to the observed formation of Deuterium.
Below we use simple electrostatic arguments to estimate
the maximal kinetic energy of accelerated protons.

The important laser parameters are related by the
formula for the laser intensity
\be
I_L = 4\hsp E_L/(\pi\hsp d_L^{\hsp 2}\hsp\tau_L ),
\ee
where $E_L$  is the total beam energy,
$\tau_L$ is the duration of the laser pulse, and
$d_L$    is the diameter of the focal spot.

For our estimates we use
the following values
(see Refs.~\cite{Szokol-NFP22,Racz-22})\hsp\footnote
{
Actually, the laser intensity has a certain
(presumably, Gaussian) radial profile
with a maximum~$I_{\rm max}$ at the beam axis.
We replace it by a~rectangular
profile with a~hight of $0.5\,I_{\rm max}$\,.
} :
\mbox{$I_L=5\cdot 10^{17}~\textrm{W/cm}^2$},
$E_L=25~\textrm{mJ}$,
\mbox{$d_L = 12.6~\mu\textrm{m}$},
the laser wave length
\mbox{$\lambda_L=0.795~\mu\textrm{m}$}, and
\mbox{$\tau_L=40~\textrm{fs}$}.

With these laser parameters
one can expect that only a
small fraction of laser energy, $\eta$,
goes to stripping of
electrons from the target's atoms.
The reason is that laser photons have relatively
low energies, only of about 1.56~eV.
Therefore, a single hydrogen atom can be ionized
if its electrons absorb at least 9 photons.


It is assumed that free electrons are accelerated
by the laser field
via the ponderomotive force proportional
to~\mbox{$\bm{j}\times\bm{B}$}.
Although this process is rapid and non-equilibrium, one can introduce
an {\it effective specific energy} of an accelerated electron, $\Theta$.
It is estimated by using the
scaling relation suggested in
~\cite{Poy2018},
assuming thermalization of absorbed laser energy:
\be
\label{Th}
\Theta\simeq 0.42\ m_e\hsp c^2
\left( 	\frac{I_L \lambda_L^2}{I_0 \lambda_0^2}  \right)^{1/3}\, .
\ee
where $m_e$ is the electron mass,  and
\mbox{$I_0 = 10^{18}\,\textrm{W/cm}^2$}, $\lambda_0 = 1\,\mu$m.

\begin{figure}[h]  
\begin{center}
\includegraphics[width=0.99\columnwidth]{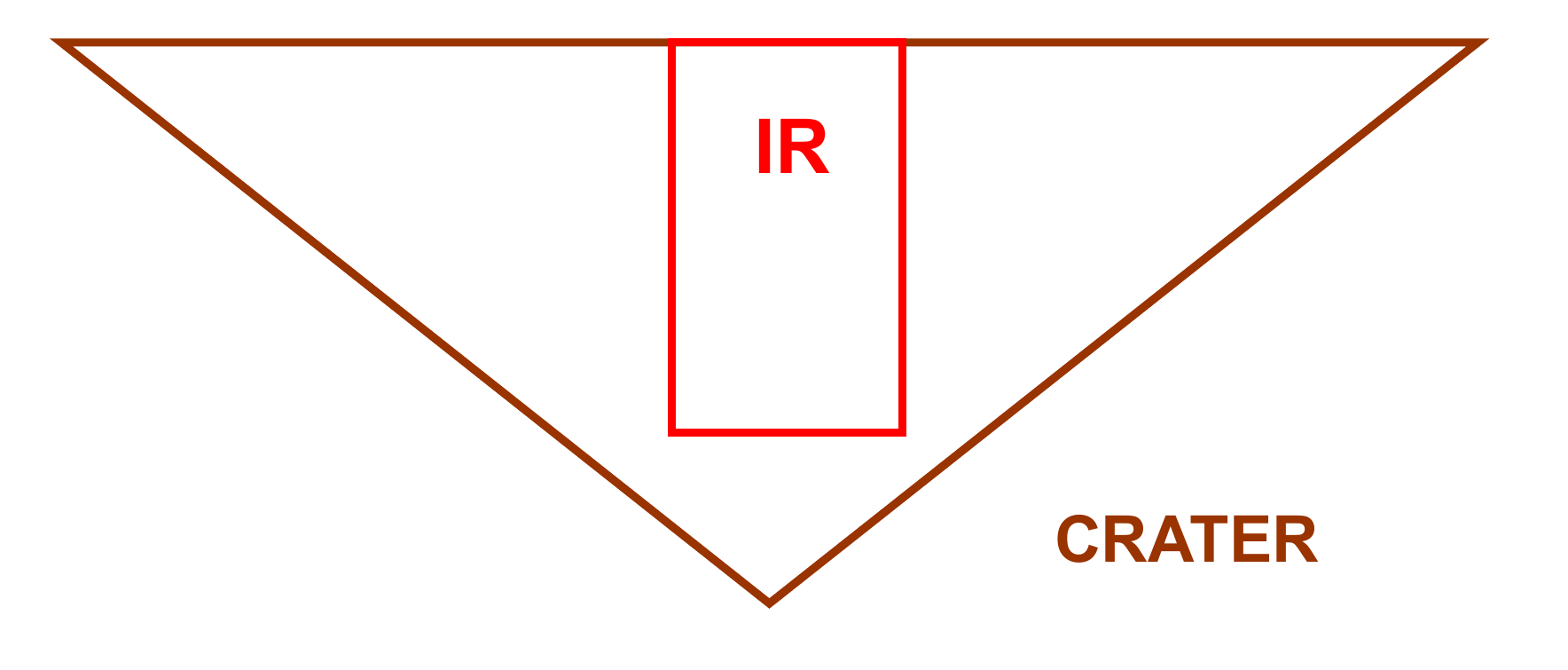}
\end{center}
\vskip -9mm
\caption{
(color online)
Schematic cross section of the final crater (C) and the laser
irradiation region (IR). The energy and momentum of the
laser beam is initially absorbed in the IR.
}
\label{F-4}
\end{figure}

We further assume that ionization electrons occupy initially the
'irradiation region'~(IR), which has approximately
a cylindrical shape with the axis along
the beam direction, as schematically shown in
Fig.~\ref{F-4}. Its diameter is
approximately equal to the laser focal spot
and its depth does not exceed the crater depth.
The initial IR is
smaller than the finally measured crater dimensions,
since part of the absorbed laser energy is spent for
heating and evaporating the surrounding matter.

Neglecting the 3D expansion effects,
one can estimate the length of IR as
\be
L_{ir}\simeq c\tau_L= 12\,\mu{\rm m}\,.
\label{lir}
\ee
The radius of this cylinder is estimated as
\be\label{rir}
R_{ir}\simeq\sqrt{\dfrac{E_L}{\pi\hsp I_L\tau_L}}\simeq
6.3\,\mu\hsp\textrm{m}\, ,
\ee
which agrees with the observed diameter of the
laser focus, $d_L=12.6\,\mu$m~\cite{Szokol-NFP22}.
The volume of the IR is
\be
V_{ir} = \pi R_{ir}^2 L_{ir} \simeq 1496\,\mu {\rm m}^3\, .
\label{vir}
\ee
In this volume the original number of H atoms in UDMA is
\be
N_{ir}^H = V_{ir} \cdot n_{U} \cdot 38 = 1.36 \cdot 10^{14} ,
\label{NHir}
\ee
where $n_{U} = 1.5 \cdot 10^{21}\,{\rm cm}^{-3}$ is the UDMA
number density
in the target.

By using the energy conservation, one can find the total number
of stripped (ionization)  electrons in the IR,
$N_{ir}^e$, as
\be\label{nae}
N_{ir}^e=\eta\,\frac{E_L}{\Theta}\,.
\ee
The parameter $\eta$ is estimated
by using the empirical formula of Ref.~\cite{Rot2017}.
\be
\eta = 0.025 \hsp \left({I_L}/{I_0}\right)^{\ 0.74}\,.
\label{eta-e}
\ee

We assume that free electrons leave quickly the IR before protons
and heavier ions start to accelerate.
Then  using the electro-neutrality, one can estimate
the net positive charge of this region as $\hsp e N_{ir}^e$
where~$e$ is the proton charge.
From Eqs.~(\ref{nae}), (\ref{eta-e}), (\ref{Th}),
one obtains the total positive charge number of~IR (in units of $e$) as
\be
N_0 \simeq N_{ir}^e \simeq 1.60 \cdot 10^{10}.
\label{npc}
\ee

This leads to the following estimate for the Coulomb potential
acting on protons
\be\label{ipot}
U\sim\frac{e\hsp N_0}{R_{ir}}\,.
\ee
Under influence of the Coulomb field, protons
(located initially inside the IR) acquire an additional energy
of the order of $e\hsp\hsp U$\,. This leads to the maximal
kinetic energy of protons
\be
E_p\sim e\hsp U\simeq 3.65~\textrm{MeV}.
\label{pme}
\ee
By substituting the above energy into Eq.~(\ref{rkin}) one obtains
that the maximal deuteron energy in the reaction~(\ref{pd})
is about 1~MeV.

At later stages of the target evolution one can expect smaller
values of acceleration due to expansion of the initially produced plasma.
It will be demonstrated in the next section that the local charge density
will significantly increase in the vicinity of NPs,
if they are imbedded in the target material.
Then one can get larger proton energies as compared
to that estimated in Eq.~(\ref{pme}). Moreover,
proton energies may even exceed the threshold values for the reaction (\ref{pd1}).

\section{The role of nano-rod antennas} 

The NAPLIFE experiments investigated  the effect
of golden NPs implanted in UDMA targets.
At 25 mJ pulse energy in the case Au1, about
\mbox{7-9 \%} Deuterium abundance was
detected~\cite{Kumari-22} compared to
the observed amount of Hydrogen atoms in the
emitted plum created after the laser shot.

The EPOCH PIC estimates~\cite{Papp2022} concluded
that a single nano-rod antenna
with dimensions \mbox{85\,nm\,$\times$\,25\, nm}
increases the electromagnetic field intensity
(in the NP's vicinity) by a factor
of 25.9 (see Eq.~(3) in
Ref.~\cite{Papp2022}).
Note however, that this result has been obtained for
relatively low intensity
\mbox{$I_L=4\cdot 10^{15}~\textrm{W/cm}^2$}.

According to these calculations, initially
about $10^3$ electrons fluctuate between the ends of the
resonant nano-antenna, with transverse peak momentum of about
$0.15$~MeV/c. The estimated maximal potential gradient is
approximately $2.9~\textrm{kV/nm}$.

Light absorption by embedded NPs was modelled
earlier at 1000--fold larger laser energy in
Ref.~\cite{Plech2006}.
However, the obtained electric field
(about $11.3$\,V/nm) is by more than two orders
of magnitude less than in our estimate. The main reason is
that non-resonant antenna/frequency combination was used!
This indicates
the importance of the use of resonant antennas. Such features are
well known for classical antennas in radio communication.

Of course, the orientation of antennas with respect to the beam direction is
rather important. Unfortunately, at present, target manufacturing
does not produce oriented nano-antenna implantation.

Recently, the proton acceleration up to~93~MeV energy
has been obtained with the PHELIX
laser at GSI~\cite{PHELIX100}.
However, proton energies exceeding 20~MeV were achieved
in these experiments only for
\mbox{$I_L>10^{20}$ W/cm$^2$}.
We expect that
employing targets with properly oriented,
resonant nano-antennas may produce
\mbox{10--20 MeV} protons even at lower intensities
\mbox{$I_L\lesssim 10^{17}$~W/cm$^2$}.

The increased energy and momentum absorption,
as well as the enhanced pulse intensity
may lead to a significant increase
of the proton acceleration in the vicinity of a~nano-rod.

Under the influence of electromagnetic field, a~single NP will be ionized,
emitting some amount of electrons~$\Delta N_s$.  Let us express the number of
ionization electrons as
\be
\Delta N_s=\xi N_{\rm Au}\,,
\ee
where $N_{\rm Au}$ is the number of atoms in a single NP, and
$\xi$ is the ratio of the ionized electrons per golden
atom\hsp\footnote
{
At $\xi\lesssim 1$ ionization electrons
come mainly from the conductivity zone of the Au metal.
}.
According to calculations of
Ref.~\cite{Kaw20},
made for laser-irradiated golden foils, one can expect the values
\mbox{$\xi\sim 30$} at laser intensities
\mbox{$I_L\gtrsim 10^{18}~\textrm{W/cm}^2$}.
We consider $\xi$ as a parameter, although it may depend on characteristics
of the laser, orientation and sizes of~NPs, etc.

A proton located near the surface of a charged nano-rod gets
an additional Coulomb energy
\be
\Delta E\sim\frac{2e^2\hsp \Delta N_s}{l_{\rm min}}
\sim 290\,\xi~\textrm{keV}\,,
\label{delE}
\ee
where $l_{\rm min}=25~\textrm{nm}$
is the diameter of the nano-rod.
In the numerical estimate we have taken
$N_{\rm Au}\simeq 2.46\cdot 10^{6}$\,.

We can also estimate the Coulomb energy per atom of the ionized NP:
\be
\frac{E_C}{N_{\rm Au}}\sim
\frac{2\,(e\Delta N_s)^2}{N_{\rm Au}l_{\rm min}}
\simeq 290\,\xi^2~\textrm{keV}\,.
\label{ECN}
\ee
If $\xi$ is not too small, this energy
will exceed the typical 'binding' energy
\mbox{$\sim 10~\textrm{eV}$} per atom in a non-excited NP.
According to Eq.~(\ref{ECN}), this occurs at
$\xi\gtrsim 6\cdot 10^{-3}$.
In this case such NPs will be destroyed
due to Coulomb-induced explosion.

It is interesting to estimate the additional
(positive) charge provided by nano-particles in the~IR.
It is equal to
$e\hsp\Delta N_s\hsp N^{(NP)}_{ir}$
where
\be
N^{(NP)}_{ir}\simeq \frac{V_{ir}}{<L_{Au2}>^3}\simeq 3850
\ee
is the number of NPs in the IR.
In the second equality we substituted
the volume of the IR, $V_{ir}$, from Eq.~(\ref{vir})
and the average distance between nano-rods, $<L_{Au2}>$,
obtained in Sec.~IIb. Again we give numerical estimates
for the case~Au2.

Due to appearance of the additional charge, protons get larger acceleration
in targets with NPs. Instead of Eqs. (\ref{ipot}) and (\ref{pme})
we obtain the following estimate for the proton energy
\be
E_p\sim \frac{e^2 (N_0+\Delta N)}{R_{ir}}
\simeq (3.65+2.15\,\xi)~\textrm{MeV},
\label{EpNR}
\ee
where
\be
\Delta N=\Delta N_s\hsp\hsp N^{(NP)}_{ir}
\simeq\hsp 9.5\cdot 10^9\,\xi
\label{DNir}
\ee
is the additional charge (in units of $e$)
of the IR due to the presence of ionized nano-rods.

At large enough $\xi$ many deuteron-production processes
with threshold energies larger than for the reaction (\ref{pd}) may be open.
For example, at $\xi\sim 10$ one gets almost 7-fold increase of proton energy,
i.e. $E_p\simeq 22~\textrm{MeV}$, which is above the
threshold energy of the reaction (\ref{pd1}).
As compared to the case without NPs, where
the number of ionized protons $N_p=N_0\lesssim 10^{-4} N^H_{ir}$
is relatively small, at $\xi\gtrsim 10$ one can expect much larger
$N_p$--values for target with nano-rods.


\section{Estimation of deuteron to proton \mbox{ratio}} 

\subsection*{a) Rate equations for deuteron production}

As discussed above, initially, ionization electrons
and protons are produced in the
cylindrical channel, which is much more narrow than the crater.
We approximate the latter
as a cone with radius $R_C$ and the depth $L_C$  (see Fig.~\ref{F-4}).
According to NAPLIFE
data the volume of the crater for case Au2 can be estimated as
\be
V_C = \frac{1}{3} \pi R_C^2 \hsp L_C \simeq
452\,400\,\mu{\rm m}^3 ,
\ee
where we have substituted the values
$R_C= 120\,\mu$m and $L_C = 30\,\mu$m\,.
One can see that $V_C$ is about 300 times larger
than the volume of the cylindrical
irradiation channel~(\ref{vir}).

The ionized protons are produced in the
cylindrical channel in a short interval
\mbox{$t\lesssim\tau_L\sim 40~\textrm{fs}$}\,.
At later times these protons propagate through the UDMA material
and may induce transmutation reactions in the
whole crater volume.
Propagation of the produced protons with
the energy \mbox{$E_p=m_p\hsp v^2/2=4~\textrm{MeV}$}
through the crater volume takes time
\be
t_C \simeq \frac{R_C}{v} =
\frac{120\,\mu\textrm{m}}{0.092\,{\rm c}} \simeq 4.35~\textrm{ps}\,,
\label{ptl}
\ee
where $m_p$ and $v$ are, respectively, the mass and velocity of the proton.

The number of deuterons, which can be produced by the
stripping reaction (\ref{pd}) can be estimated
by using a simple rate equation
\be
\frac{dN_d}{dt} = \langle \sigma_s v \rangle\, n_{13C} \cdot  n_p(t) V(t) \, ,
\label{rate}
\ee
where $\sigma_s$ is the cross section of the reaction~(\ref{pd}),
$n_{13C}\simeq\textrm{const}$ is the number density of $^{13}C$ in the
UDMA matter, and $V(t)$ is the volume occupied by the ionization protons.

Obviously, after a short initial stage, \mbox{$t>\tau_L$}, the total number
of protons remains the same, but their density drops, such that
$N_p=n_p\hsp V(t)=\textrm{const}$, where $N_p$ is given by Eq.~(\ref{npc}).
Therefore, all terms in the right hand side of Eq.~(\ref{rate}) are
approximately time-independent and one may write its solution
at $t>\tau_L$ as
\be
\frac{N_d(t)}{N_p} = \langle \sigma_s v \rangle\, n_{13C}
\cdot (t-\tau_L)
\label{tdep}
\ee
Now we can estimate the $d/p$ ratio taking \mbox{$t-\tau_L\sim t_C$},
choosing the input parameters: $\sigma_s\simeq 2~\textrm{mb}$,
$v\simeq 0.092\,c\simeq \mbox{$2.76 \cdot 10^9\,{\rm cm}/s$}$ (at $E_p\simeq 4~\textrm{MeV}$)
and substituting \mbox{$n_{13C} = 3.66\cdot 10^{20}\,{\rm cm}^{-3}$}.
Finally we get the following ratio (for the case Au2)
\ba
\frac{N_d}{N_p} &=&
5.52 \cdot 10^{-18} \frac{\rm cm^3}{s}\,\times
3.66 \cdot 10^{\,20} \frac{1}{\rm cm^3}\,\nonumber\\
&\times& 4.35 \cdot 10^{-12} s\simeq 8.8 \cdot 10^{-9}.
\label{dpr1}
\ea

Now let us estimate the $d/p$ radio for the reaction (\ref{pd1}).
One can do this using the Eqs.~(\ref{rate})
and (\ref{tdep}), but replacing $n_{13C}$ by the two-order of magnitude larger density
of \mbox{$^{12\hsp}\textrm{C}$} nuclei \mbox{$n_{12C}\simeq 3.41\cdot 10^{22}~\textrm{cm}^{-3}$},
and substituting the cross section $\sigma_s$ of the reaction (\ref{pd1}).
For a rough estimate we choose the proton energy $E_p=20~\textrm{MeV}$ and the cross section
$\sigma_s\simeq 25~\textrm{mb}$ from Ref.~\cite{Whi58}.
Then, instead of~(\ref{dpr1}) one obtains the estimate
\ba
\frac{N_d}{N_p} &=&
1.55 \cdot 10^{-16} \frac{\rm cm^3}{s}\,\times
3.41 \cdot 10^{\,22} \frac{1}{\rm cm^3}\,\nonumber\\
&\times& 1.94 \cdot 10^{-12} s\simeq 1.02 \cdot 10^{-5},
\label{dpr2}
\ea
which is by a factor 1160 larger than the $d/p$ ratio
for the reaction (\ref{pd}).

Nevertheless, both $d/p$ values are too small to explain the D/H ratios
extracted from the LIBS spectra~\cite{Racz-22,Kumari-22}.
Note that both ratios are below the background ratio
$ \frac{N_D}{N_H}=1.6 \cdot 10^{-4}$
in the Earth atmosphere under normal conditions.

However, one should bear in mind that
to be visible in LIBS spectra, protons and deuterons
should form
neutral H and D atoms by recombination with electrons
in the UDMA target.
Therefore, the LIBS measurements give only indirect
information about the abundances of ionized
deuterons and protons, produced by the irradiation of UDMA targets.
It is clear that the recombination process will shift
the D/H ratio to larger values. The reason is that deuterons produced in
reactions (\ref{pd}), (\ref{pd1}) have significantly lower velocities as
compared to protons. But, the recombination cross section increases
with decreasing particle velocity. Therefore,
the recombination probability for deuterons will be larger then for protons.

\subsection*{b) Role of electron-ion recombination}

To estimate the numbers of the atomic hydrogen (H) and Deuterium (D)
one can use the rate equations similar to (\ref{rate}), but replacing
$\sigma_s$ by
the recombination cross section $\sigma_r$  and $n_{13C}$ by
the electron density $n_e$.
Then the ratio D/H will be given by the equation similar to (\ref{tdep})
with the replacements $n_{13C}\to n_e$ and $\sigma_s\to\sigma_r$.

To get the quantitative estimate, we assume that recombination lengths
for protons and deuterons are longer than the crater size
$\sim 100~\mu$m\hsp .
Then the ratio of recombination probabilities will be approximately equal
to the ratio of recombination cross sections. According to
Ref.~\cite{Kot2019}, they are inversely proportional to the fifth
power of the ion velocity~\cite{elve}.
Finally, we obtain the approximate relation
\be
\frac{\rm D}{\rm H}\sim
\left(\frac{m_d\hsp E_p}{m_p\hsp E_d}\right)^{5/2}\frac{d}{p}\,.
\label{dhe}
\ee

Note that we have assumed that propagation times of proton and deuterons
in the target are inversely proportional to ion velocities.
Now we can calculate the D/H ratios for two
stripping reactions (\ref{pd}) and (\ref{pd1}) considered before.

For the reaction (\ref{pd}), taking
$E_p=4~\textrm{MeV}$ and $E_d=1~\textrm{MeV}$,
one has from Eqs.~(\ref{dpr1}), (\ref{dhe})
\be
\frac{\rm D}{\rm H}\sim 181\times\frac{d}{p}\simeq 1.6 \cdot 10^{-6}.
\label{dh4}
\ee
One can see that this reaction still gives too small ${\rm D}/{\rm H}$ ratio
even if one takes into account the recombination corrections.

In the case of the reaction (\ref{pd1}), substituting
\mbox{$E_p=20~\textrm{MeV}$},
$E_d=5.92~\textrm{MeV}$ (this value follows from Eq.~(\ref{rkin})), and using
Eqs.~(\ref{dpr2}), (\ref{dhe}), one gets the estimate
\be
\frac{\rm D}{\rm H}\sim 118\times\frac{d}{p}\simeq 1.2 \cdot 10^{-3}.
\label{dh4}
\ee
This value is still below experimental ratios
for the Au2 target, but it is much closer to them. We plan to check
these results by using more accurate calculations in the future.
In particular, we are going to take into account feeding of the
Balmer-$\alpha$ states from higher atomic levels.

\section{Conclusion and Outlook}

In the present article we have analysed the results
of NAPLIFE experiments on laser irradiation
of polymer targets. They show a significant formation
of deuteron atoms at high enough laser pulse
energies. The deuteron's yield increases with the
concentration of implanted nano-particles.

In our analysis we have focused on the possibility of
nuclear transmutation mechanisms,
namely, due to the stripping reactions
\mbox{$p+\textrm{C}\to d+X$} with the carbon
isotopes $^{12\hsp}\textrm{C}$ and $^{13\hsp}\textrm{C}$.
It is argued that proton energies achievable
in the laser--target interaction are sufficient
for deuteron formation in such reactions, especially
for targets with implanted golden NPs.

Additional investigations are necessary for verifying
these results.
First, it would be desirable to reproduce our observations
at higher laser energies e.g. at GSI/FAIR PHELIX
and future ELI-ALPS facilities. Additional diagnostics
tools would be desirable, including measurements
of X-rays, neutrons, $\alpha$-particles, etc.
We plan to perform new experiments with different
pulse widths, target thickness and profiles of
NPs.

On the theoretical side, the numerical modeling of
target plasma evolution is necessary to estimate yields
of secondary particles and their momentum distributions.
Special investigation should be made to study
properties of hot-spots created by nano-particles,
and their role in the nuclear transmutation.



\section*{Acknowledgments}

Enlightening discussions with Mikl\'os Kedves, M\'ark Aladi, and
Oliver A. Fekete are gratefully acknowledged.
T.S. Bir\'o, M. Csete, N. Kro\'o, I. Papp, A. Szenes, and D. Vass
acknowledges support by the National Research, Development and
Innovation Office (NKFIH) of Hungary.
Horst St\"ocker acknow\-ledges the Judah M. Eisenberg Professor Laureatus
chair at Fachbereich Physik of Goethe Universit\"at Frankfurt.
D\'enes Moln\'ar acknowledges support by the US Department of Energy,
Office of Science, under Award No. DE-SC0016524.
We would like to thank the Wigner GPU Laboratory at the
Wigner Research Center for Physics for providing support
in computational resources.
This work is supported in part by
the Frankfurt Institute for Advanced Studies, Germany,
the E\"otv\"os, Lor\'and Research Network of Hungary,
the Research Council of Norway, grant no. 255253, and
the National Research, Development and Innovation Office of Hungary,
for projects:
Nanoplasmonic Laser Fusion Research Laboratory under
project Nr-s NKFIH-874-2/2020 and NKFIH-468-3/2021,
Optimized nanoplasmonics (K116362), and
\mbox{Ultrafast} physical processes in atoms, molecules,
nanostructures and biological systems (EFOP-3.6.2-16-2017-00005).


\clearpage


\begin{thebibliography}{99}
\bibitem{elcon}
Note, that \mbox{$p\to n$} conversion may also occur
by capturing electrons from bound atomic states.
However, such processes are slow and may occur
only for sufficiently heavy atoms.
\bibitem{pptod}
One can also consider the deuteron
production in exothermic reactions
\mbox{$p+p\to d+e^++\overline{\nu}_e$}.
But this
weak reaction has nonzero Coulomb barrier
and the corresponding cross section is relatively small.
\bibitem{elve}
For a rough estimate we assume that target's electrons have much smaller velocities than ions.
\end{thebibliography}
\end{document}